# On the incorrectness of the proof of the Gibbs theorem on the entropy of a mixture of ideal gases, which was given by J. W. Gibbs [1]

## V. Ignatovych

Department of Philosophy, National Technical University of Ukraine «Igor Sikorsky Kyiv Polytechnic Institute»

e-mail: v.ihnatovych@kpi.ua

**Abstract**

There is the Gibbs theorem in thermodynamics, according to which the entropy of the mixture of ideal gases is equal to the sum of the entropies of the components of the mixture. J. W. Gibbs proved this by a mathematical derivation from the beginnings of thermodynamics and the laws of ideal gases, in particular Dalton's law. This message shows that this proof is mathematically incorrect. It is also shown that the Gibbs theorem cannot be derived mathematically from Dalton's law.

### 1. Introduction

Gibbs' theorem on the entropy of a mixture of ideal gases expressed by the formula:

$$\eta_c = \sum_i \eta_i \qquad (I)$$

where $\eta_c$ is the entropy of the mixture of ideal gases, $\eta_i$ is the entropy of the i-th ideal gas-component of the mixture.

The proof of Gibbs' theorem, which was given by J. W. Gibbs, is published in his paper «On the Equilibrium of Heterogeneous Substances» [1, c. 55-353]. It consists in deriving a formula expressing the Gibbs theorem from other formulas.

Dalton's law according to which the pressure of a mixture of ideal gases is equal to the sum of the partial pressures of individual gases is used in this proof:

$$p = \sum_i p_i. \qquad (II)$$

Let us analyze this proof.

---

[1] The fragments of the article "On the incorrectness of the proofs of the Gibbs theorem on the entropy of a mixture of ideal gases" [In Russian] // Proceeding of Sixth International Scientific-Practical Conference "Mathematics in Modern Technical University", Kyiv, December, 28–29, 2017. Kyiv: Igor Sikorsky Kyiv Polytechnic Institute, 2018. P.324-330. URL: http://matan.kpi.ua/public/files/2017/mvstu6/MSTU6.pdf.

## 2. An analysis of the proof of the Gibbs theorem, which Gibbs gave

On p. 63 Gibbs writes down expressions for the total differential of the internal energy $\varepsilon$ of a homogeneous thermodynamic system (in his terminology - the homogeneous mass) as a function of the extensive parameters of the system:

$$d\varepsilon = td\eta - pdv \tag{11}$$

$$d\varepsilon = td\eta - pdv + \mu_1 dm_1 + \mu_2 dm_2 + ... + \mu_n dm_n, \tag{12) (86}$$

where $t$, $p$, $\mu_1$, $\mu_2$ ..., $\mu_n$ — are partial derivatives $\varepsilon$ with respect to $\eta$, $v$, $m_1$, $m_2$, ..., $m_n$; $t$ — temperature, $\eta$ — entropy, $p$ — pressure, $v$ — volume, $\mu_i$ — the chemical potential of the i-th substance, $m_i$ — is the mass of the i-th substance.

Hereinafter, the numbering of formulas by Arabic numerals is given by the work of Gibbs. Formulas given by the author and unnumbered formulas from the work of Gibbs enumerate by Roman numerals.

Gibbs points out (ibid.) that the formula (11) refers to the case when the amount and kind of matter in this mass (system) as fixed, and (12) — to the case when the matter in the mass (system) as variable. He repeats (12) on p. 86 under the number (86).

Further, Gibbs defines a series of functions ($\psi$, $\chi$, $\zeta$) and, taking (86) into account, derives formulas for their total differentials. On p. 87 he records:

$$\zeta = \varepsilon - t\eta + pv, \tag{91}$$

$$d\zeta = -\eta dt + vdp + \mu_1 dm_1 + \mu_2 dm_2 + ... + \mu_n dm_n. \tag{92}$$

He further writes (p. 87):

«If we integrate (86), supposing the quantity of the compound substance considered to vary from zero to any finite value, its nature and state remaining unchanged, we obtain

$$\varepsilon = t\eta - pv + \mu_1 m_1 + \mu_2 m_2 + ... + \mu_n m_n \tag{93}$$»

Further he obtains similar formulas for $\psi$, $\chi$, $\zeta$. To this place, the editor of the book gave the following comment: «A rigorous justification for this integration procedure is usually made with reference to Euler's theorem on homogeneous functions ...» [2, p.516]. Let us explain what this means.

Gibbs pointed out: «We know… *a priori,* that if the quantity of any homogeneous mass containing $n$ independently variable components varies and not its nature or state, the quantities $\varepsilon$, $\eta$, $v$, $m_1$, $m_2$, ..., $m_n$ will all vary in the same proportion» [ibid., p. 86].

Gibbs considered the equilibrium in systems of reactive substances. The nature and state of the substances of the system does not change with constant values of temperature and pressure. The words «the quantities $\varepsilon$, $\eta$, $v$, $m_1$, $m_2$, …, $m_n$ will all vary in the same proportion» with respect to $\varepsilon$ mean that $\varepsilon$ is a homogeneous function of the first power of the quantities $\eta$, $v$, $m_i$:

$$\varepsilon(a\eta, av, am_i) = a\varepsilon(\eta, v, m_i).\tag{V}$$

The total differential of the internal energy of a system of substances has the form:

$$d\varepsilon = \frac{\partial \varepsilon}{\partial \eta}d\eta - \frac{\partial \varepsilon}{\partial v}dv + \frac{\partial \varepsilon}{\partial m_1}dm_1 + \frac{\partial \varepsilon}{\partial m_2}dm_2 + ... + \frac{\partial \varepsilon}{\partial m_n}dm_n.\tag{VI}$$

If $\varepsilon$ is a homogeneous function of the first power of the quantities $\eta$, $v$, $m_i$, then, on the basis of Euler's theorem on homogeneous functions [3, p. 108], from (VI) we can derive a formula equivalent to the formula (93):

$$\varepsilon = \frac{\partial \varepsilon}{\partial \eta}\eta - \frac{\partial \varepsilon}{\partial v}v + \frac{\partial \varepsilon}{\partial m_1}m_1 + \frac{\partial \varepsilon}{\partial m_2}m_2 + ... + \frac{\partial \varepsilon}{\partial m_n}m_n.\tag{VII}$$

We draw attention to the fact that formula (93) is obtained by integrating (86) under the condition that $p$, $t$, $\mu_i$ do not change. However, then Gibbs differentiates the formula (93) as if all the quantities on the right are independent variables, and, comparing the result with (86), obtains (p.88)

$$-vdp + \eta dt + m_1 d\mu_1 + m_2 d\mu_2 + ... + m_n d\mu_n = 0,\tag{97}$$

and from it the next formula derives:

$$dp = \frac{\eta}{v}dt + \frac{m_1}{v}d\mu_1 + \frac{m_2}{v}d\mu_2 + ... + \frac{m_n}{v}d\mu_n.\tag{98}$$

Gibbs used (98) in the proof of the Gibbs theorem (pp. 155-156) by deriving a formula expressing the entropy of a mixture $\eta$ through entropy of components as functions $t$, $v$, $m_i$.

Let us consider in detail his calculations.

On p. 150 Gibbs gives the equations for the unit mass of an ideal gas:

$$pv = at,\tag{V}$$

$$d\varepsilon = cdt,\tag{VI}$$

where $a$ they $c$ are also constants.

Integrating the second equation, Gibbs obtains:

$$\varepsilon = ct + E.\tag{VII}$$

From these equations and (11) Gibbs obtains the formula:

$$d\varepsilon = \frac{\varepsilon - E}{c}d\eta - \frac{a}{v}\frac{\varepsilon - E}{c}dv.\tag{VIII}$$

He converts (VIII) as follows:

$$c\frac{d\varepsilon}{\varepsilon - E} = d\eta - a\frac{dv}{v}, \qquad (IX)$$

and then integrates:

$$c\ln\frac{\varepsilon - E}{c} = \eta - a\ln v - H. \qquad (X)$$

Further (p. 150) Gibbs writes: «We may extend the application of the equation to any quantity of the gas, without altering the values of the constants, if we substitute $\frac{\varepsilon}{m}$, $\frac{\eta}{m}$, $\frac{v}{m}$ for $\varepsilon$, $\eta$, $v$ respectively. This will give

$$c\ln\frac{\varepsilon - Em}{cm} = \frac{\eta}{m} - H + a\ln\frac{m}{v}. \qquad (255)»$$

The transition from (X) to (255) is based on the abovementioned provision that $\varepsilon$ is a homogeneous function of the first power of quantities $\eta$, $v$, $m_i$. Differentiating (255) and comparing the result with (86), Gibbs obtained formulas (257) - (259) for $t$, $p$, $\mu$ (p.151). Then he outputs a series of formulas for $\psi$, $\xi$, $\eta$, $p$, $\mu$, including the following:

$$\eta = m\left(H + c\ln t + a\ln\frac{v}{m}\right). \qquad (262)$$

$$\xi = Em + mt\left[c + a - H - (c+a)\ln t + a\ln\frac{p}{a}\right]. \qquad (265)$$

Differentiating (265) and comparing with (92), Gibbs obtains the formulas (p. 152):

$$\eta = m\left[H + (c+a)\ln t - a\ln\frac{p}{a}\right], \qquad (266)$$

$$\mu = E + t\left[c + a - H - (c+a)\ln t + a\ln\frac{p}{a}\right] \qquad (268)$$

From (268) Gibbs deduces the formula (p. 152):

$$p = ae^{\frac{H-c-a}{a}} t^{\frac{c+a}{a}} e^{\frac{\mu-E}{at}}. \qquad (270)$$

From (270), on the basis of Dalton's law, Gibbs obtains an expression for the pressure of a mixture of ideal gases (p.155):

$$p = \sum_1 \left(a_1 e^{\frac{H_1-c_1-a_1}{a_1}} t^{\frac{c_1+a_1}{a_1}} e^{\frac{\mu_1-E_1}{a_1 t}}\right). \qquad (273)$$

Differentiating (273) and comparing with (98), Gibbs obtains (p.156):

$$\frac{\eta}{V} = \sum_1 \left( \left( c_1 + a_1 - \frac{\mu_1 - E_1}{t} \right) e^{\frac{H_1 - c_1 - a_1}{a_1}} t^{\frac{c_1}{a_1}} e^{\frac{\mu_1 - E_1}{a_1 t}} \right), \tag{274}$$

$$\left.\begin{aligned} \frac{m_1}{v} &= e^{\frac{H_1 - c_1 - a_1}{a_1}} t^{\frac{c_1}{a_1}} e^{\frac{\mu_1 - E_1}{a_1 t}}, \\ \frac{m_2}{v} &= e^{\frac{H_2 - c_2 - a_2}{a_2}} t^{\frac{c_2}{a_2}} e^{\frac{\mu_2 - E_2}{a_2 t}}, \end{aligned}\right\} \tag{275}$$

etc.

since, according to (98), $\partial p / \partial t = \eta / v$, $\partial p / \partial \mu_i = m_i / v$.

From (275), Gibbs deduces (276) for $\mu_i$, excludes with its help $\mu_i$ from (274), and obtains the formula that is a variant of the Gibbs theorem (I):

$$\eta = \sum_1 \left( m_i H_i + m_i c_i \ln t + m_i a_i \ln \frac{v}{m_i} \right). \tag{278}$$

The derivation of the formula (I) from (274) and (275) can be done in a different way. It follows from (266) and (268) that in (274) the first factor in parentheses is equal to $\eta_i$. The product of the last three factors of (274), according to (275), is equal to $m_i / V$. This implies (I).

### 3. Incorrectness of the Gibbs proof.

In the proof of Gibbs, the following consequences of the formula (98): $\partial p / \partial t = \eta / v$, $\partial p / \partial \mu_i = m_i / v$ are very important.

However, Gibbs overlooked the fact that the formula (98) is a consequence of (93). The formula (93) is obtained from the formula (86) under the condition that $p$, $t$, $\mu_i$ do not change. Accordingly, and the conclusions $\partial p / \partial t = \eta / v$, $\partial p / \partial \mu_i = m_i / v$ in the formula (98) $dp = 0$, $dt = 0$, $d\mu_i = 0$ are incorrect.

### 4. The role of incorrect formulas in the proof of Gibbs.

The formula (I) means that entropy is an additive property of a mixture of ideal gases: the entropy of the whole (mixture) is equal to the sum of the entropies of the parts of the mixture (individual gases). The formula (II) means that the pressure is also an additive property of a mixture of ideal gases. Since entropy is a function of pressure, for a mixture of two gases from (I) and (II) follows:

$$S(p_1 + p_2) = S(p_1) + S(p_2). \tag{XI}$$

The formula (XI) means that entropy is an additive function of pressure. This is possible only if the entropy is a homogeneous linear function of pressure, as only homogeneous linear function are additive (see, E.g., [4, pp. 157-158]).

According to (266), the entropy is a logarithmic function of the gas pressure, respectively, is not an additive pressure function. It does not follow from (266) that $\partial p / \partial t = \eta / v$. If the entropy of ideal gases is expressed by (266), then the conclusion that the entropy of a mixture of ideal gases is equal to the sum of the entropies of the components of the mixture cannot be derived on the basis of Dalton's law.

**5. Conclusion.**

The proof of Gibbs' theorem, which Gibbs gave, is incorrect. The conclusion that the entropy of a mixture of ideal gases is equal to the sum of the entropies of the components of the mixture cannot be derived mathematically from the Dalton law.

**Thanks**

The author expresses his gratitude to associate professor V. Haidey and professor V. Gorshkov for useful comments.